\documentclass[twocolumn,floatfix,superscriptaddress,a4paper,showpacs,showkeys,nofootinbib,reprint,prd]{revtex4-1}
\usepackage{epsfig}
\usepackage{latexsym}
\usepackage{xspace}
\usepackage{hyperref}
\usepackage[latin2]{inputenc}
\usepackage{indentfirst}
\usepackage{enumerate}
\usepackage{color}

\usepackage{amsmath}
\usepackage{amssymb}
\usepackage[english]{babel}
\usepackage{url}
\newcommand{\eq}[1]{\begin{align} #1 \end{align}}

\begin{document}

%TC:ignore

\title{
Repulsive baryonic interactions and lattice QCD observables\\
at imaginary chemical potential
}

\author{Volodymyr Vovchenko}
\affiliation{
Institut f\"ur Theoretische Physik,
Goethe Universit\"at Frankfurt, Max-von-Laue-Str. 1, D-60438 Frankfurt am Main, Germany}
\affiliation{Frankfurt Institute for Advanced Studies, Giersch Science Center, Goethe Universit\"at Frankfurt, Ruth-Moufang-Str. 1, D-60438 Frankfurt am Main, Germany}
\affiliation{
Department of Physics, Taras Shevchenko National University of Kiev, Glushkova Ave 2, 03022 Kiev, Ukraine}

\author{Attila P\'asztor}
\affiliation{
Department of Physics, Wuppertal University, Gaussstr. 20, D-42119 Wuppertal, Germany}

\author{Zolt\'an Fodor}
\affiliation{
Department of Physics, Wuppertal University, Gaussstr. 20, D-42119 Wuppertal, Germany}
\affiliation{
Institute for Theoretical Physics, E\"otv\"os University,
P\'azm\'any P. s\'et\'any 1/A, H-1117 Budapest, Hungary}
\affiliation{
J\"ulich Supercomputing Centre, Forschungszentrum J\"ulich, D-52425 J\"ulich, Germany}

\author{Sandor D. Katz}
\affiliation{
Institute for Theoretical Physics, E\"otv\"os University,
P\'azm\'any P. s\'et\'any 1/A, H-1117 Budapest, Hungary}
\affiliation{
MTA-ELTE "Lend\"ulet" Lattice Gauge Theory Research Group,
P\'azm\'any P. s\'et\'any 1/A, H-1117 Budapest, Hungary}

\author{Horst Stoecker}
\affiliation{
Institut f\"ur Theoretische Physik,
Goethe Universit\"at Frankfurt, Max-von-Laue-Str. 1, D-60438 Frankfurt am Main, Germany}
\affiliation{Frankfurt Institute for Advanced Studies, Giersch Science Center, Goethe Universit\"at Frankfurt, Ruth-Moufang-Str. 1, D-60438 Frankfurt am Main, Germany}
\affiliation{
GSI Helmholtzzentrum f\"ur Schwerionenforschung GmbH, Planckstr. 1, D-64291 Darmstadt, Germany}

\begin{abstract}
The first principle lattice QCD methods allow to calculate the thermodynamic observables at finite temperature and imaginary chemical potential. These can be compared to the predictions of various phenomenological models. 
We argue that Fourier coefficients with respect to imaginary baryochemical potential are sensitive
to modeling of baryonic interactions. As a first application of this sensitivity, we
consider the hadron resonance gas (HRG) model with repulsive baryonic interactions, which 
are modeled by means of the excluded volume correction.
The Fourier coefficients of the imaginary part of the net-baryon density at imaginary baryochemical potential -- corresponding to the fugacity or virial expansion at real chemical potential -- are calculated within this model, and compared with the $N_t = 12$ lattice data.
The lattice QCD behavior of the first four Fourier coefficients up to $T \simeq 185$~MeV is described fairly well by an interacting HRG with a single baryon-baryon eigenvolume interaction parameter $b \simeq 1$~fm$^3$, while
the available lattice data on the difference $\chi_2^B - \chi_4^B$ of baryon number susceptibilities is reproduced up to $T \simeq 175$~MeV.

\end{abstract}



\maketitle

%TC:endignore 

\section{Introduction}

The Monte Carlo lattice QCD simulations provide the equation of state of the (2+1)-flavor strongly interacting matter at zero chemical potential~\cite{Borsanyi:2013bia,Bazavov:2014pvz}.
A crossover-type transition is observed~\cite{Aoki:2006we}.
The pseudocritical temperature $T_{pc}$ of the transition depends on the observable used to define it, estimates based on chiral condensate and its susceptibility give $T_{pc} \approx 155$~MeV~\cite{Borsanyi:2010bp,Bazavov:2011nk}, while observables based on strangeness suggest somewhat higher temperatures~\cite{Borsanyi:2010bp,Bellwied:2013cta}.
Below the transition one expects to find the confined hadronic phase. Many lattice QCD observables in that temperature range are indeed well described by a simple ideal hadron resonance gas (HRG) model ~\cite{Borsanyi:2011sw,Bazavov:2012jq,Bellwied:2015lba,Bellwied:2013cta}. 

It was pointed out recently, that the behavior of lattice observables in the crossover region, particularly of correlations and fluctuations of conserved charges, is very sensitive to the modeling of the baryonic interactions~\cite{Vovchenko:2016rkn,Vovchenko:2017cbu}.
This sensitivity is of great interest, since hadronic modeling of conserved charge fluctuations is often used to extract 
freeze-out parameters of heavy ion collisions~\cite{Alba:2014eba,Alba:2015iva}.
Lattice observables at finite net baryon density can certainly be expected to be even more sensitive to the modeling of these interactions. Unfortunately, direct Monte Carlo calculations at finite $\mu_B$ are hindered by the sign problem.
Main methods to circumvent this problem include the reweighing techniques~\cite{Barbour:1997ej,Fodor:2001au,Fodor:2001pe,Csikor:2002ic}, the Taylor expansion around $\mu = 0$~\cite{Allton:2002zi,Allton:2005gk,Gavai:2008zr,Basak:2009uv}, and the analytic continuation from imaginary $\mu$~\cite{deForcrand:2002hgr,DElia:2002tig,Wu:2006su,deForcrand:2008vr,DElia:2009pdy,Philipsen:2014rpa,Cuteri:2015qkq,DElia:2016jqh,Alba:2017mqu}.
These methods have allowed to calculate some thermodynamic features of QCD at small but finite chemical potentials~\cite{Bellwied:2015rza,Gunther:2016vcp,Bazavov:2017dus}.

In the present work we consider the imaginary $\mu$ method. 
We use the updated version of the lattice data, shown previously in Ref.~\cite{Borsanyi:QM2017}.
However, instead of performing analytic continuation from imaginary chemical potential to real chemical potential,
we instead directly compare lattice data at imaginary $\mu$ to the corresponding predictions of the phenomenological models.
Some phenomenological models were considered at imaginary chemical potential before, such as the quasiparticle model~\cite{Bluhm:2007cp} or the PQM model~\cite{Morita:2011jva}.
In the present work, our focus is on the HRG model with repulsive interactions for baryon-baryon and antibaryon-antibaryon pairs, modeled by means of the excluded volume (EV) correction.

The paper is organized as follows: in Sec.~\ref{sec:QCDobs} the lattice observables at imaginary baryochemical potential, which are studied in the present work, are introduced. 
Sec.~\ref{sec:models} lists the predictions for these observables from several phenomenological models.
The lattice method is described in Sec.~\ref{sec:lattice},
and in Sec.~\ref{sec:results} lattice results are compared to the predictions of interacting HRG models.
Summary in Sec.~\ref{sec:summary} closes the article.

\section{QCD observables at imaginary baryochemical potential}
\label{sec:QCDobs}

Due to the baryon-antibaryon symmetry, the QCD pressure is an even function of a real baryochemical potential 
$\mu_B$
at a finite temperature. This quantity can then be written as the following series expansion:
\eq{\label{eq:pseries}
\frac{p(T,\mu_B)}{T^4} 
= \sum_{k = 0}^{\infty} \, p_k (T) \, \cosh( k \, \mu_B / T ),
}
provided that the expansion is convergent at a given $T$-$\mu_B$ pair\footnote{Throughout this work we assume that strangeness and electric charge chemical potentials are zero, i.e. $\mu_S = \mu_Q = 0$.}. 
At $\mu_B=0$, the pressure is simply the sum of all coefficients $p_k(T)$. Therefore these can be interpreted as the partial pressures,
coming from the sectors of the Hilbert space with a different baryon number.

The first-order net baryon susceptibility 
$\chi_1^B(T, \mu_B) \equiv \partial (p/T^4) / \partial (\mu_B / T)$ 
is proportional to the net baryon density and it is equal to
\eq{\label{eq:chi1def}
\chi_1^B(T, \mu_B) = \frac{\rho_B (T, \mu_B)}{T^3} = \sum_{k = 1}^{\infty} \, b_k (T) \, \sinh( k \, \mu_B / T ),
}
where, by definition,
\eq{
b_k (T) \equiv k \, p_k(T).
}
It is clear that the knowledge of all $a_k(T)$ coefficients provides complete information about the thermodynamic properties of QCD in the region of the phase diagram where the series expansion given by Eq.~\eqref{eq:pseries} is convergent.

One can consider the susceptibility $\chi_1^B$ in Eq.~\eqref{eq:chi1def} at a purely imaginary value of the baryochemical potential, i.e. at $\mu_B = i \, \tilde{\mu}_B$.
The analytic continuation yields
\eq{
\label{eq:chi1imag}
\chi_1^B(T, i \tilde{\mu}_B) = i \sum_{k = 1}^{\infty} \, b_k (T) \, \sin( k \, \tilde{\mu}_B / T ),
}
i.e. the $\chi_1^B$ itself becomes purely imaginary.
The imaginary part of $\chi_1^B$ in Eq.~\eqref{eq:chi1imag} has explicit form of the trigonometric series expansion, with $b_k(T)$ being the corresponding temperature dependent Fourier coefficients.
If the $\tilde{\mu}_B$-dependence of $\chi_1^B$ is known (e.g. from lattice simulations), then the coefficients $b_k(T)$ can be calculated in the standard way:
\eq{\label{eq:bFourier}
b_k(T) = \frac{2}{\pi} \int_0^{\pi} \, d \tilde{\mu}_B \, [\textrm{Im} \, \chi_1^B(T, i \tilde{\mu}_B)] \, \sin( k \, \tilde{\mu}_B / T ) .
}

\section{Phenomenological models}
\label{sec:models}

In some analytic models of the equation of state, the coefficients $b_k(T)$ can be worked out explicitly.

\subsection{Ideal HRG}

A popular model to describe the confined phase of QCD at low temperatures is the hadron resonance gas model.
In its simplest implementation,
the system is modeled as a non-interacting mixture of all known hadrons and resonances. 
It is argued~\cite{Dashen:1969ep}, that the
inclusion into the model of all known resonances as free non-interacting (point-like) particles, may allow for an
effective modeling of the attractive interactions between hadrons, including 
the formation of narrow resonances and of Hagedorn states.
This ideal HRG model 
has a long history of being used to describe the hadron production in heavy-ion collisions at various collision energies~\cite{Cleymans:1992zc,Cleymans:1998fq,Becattini:2003wp,Andronic:2005yp,Letessier:2005qe}.

In the present HRG analysis we employ the Boltzmann approximation for all baryons. This is a good approximation for the observables of interest.
We do not include the light nuclei into the HRG particle list.
The inclusion of nuclei would induce nonzero $b_2,\,b_3,\ldots$, but always with a positive sign. This is 
in contrast to our lattice results, e.g. that $b_2<0$, indicating that the next important correction to the HRG model is not from these states, but from repulsive interactions.
The net baryon density $\rho_B^{\rm id}$ in the ideal HRG model reads
\eq{\label{eq:idHRG}
\rho_B^{\rm id} (T, \mu_B) = 2 \, \phi_B(T) \, \sinh(\mu_B / T),
}
where
\eq{
\phi_B (T) = \sum_{i \in B} \, \int d m \, \rho_i(m) \, \frac{d_i \, m^2 \, T}{2\pi^2} \, K_2\left( m \over T \right)
}
is the baryonic spectrum,
with $d_i$ and $\rho_i$ being, respectively, the degeneracy and a properly normalized mass distribution for hadron type $i$, and where the sum goes over all baryons in the system. 
Note that the summation does not include antibaryons.
We include the baryon states, which are listed in the Particles Data Tables~\cite{Olive:2016xmw} and have a confirmed status there.
The function $\rho_i$ takes into account the non-zero widths of the resonances by the additional integration over their Breit-Wigner shapes, following Refs.~\cite{Becattini:1995if,Wheaton:2004qb}.

It is evident from Eq.~\eqref{eq:idHRG} that all Fourier coefficients $b_k^{\rm id}$ are equal to zero for $k \geq 2$. For the first coefficient one obtains $b_1^{\rm id} (T) = 2 \, \phi_B(T) / T^3$.

\subsection{HRG with repulsive baryonic interactions}
\label{sec:EVHRG}

In a more realistic HRG model one has to also take into account the attractive and repulsive interactions between hadrons which cannot be attributed to the resonance formation.
In particular, the nucleon-nucleon interaction is known to be largely repulsive at short distances and the corresponding scattering phase shifts are not known to exhibit any resonance structure.
The importance of the van der Waals like interactions between baryons for lattice QCD observables was recently pointed out in Ref.~\cite{Vovchenko:2016rkn}.
In the present work we perform similar analysis for the observables at imaginary chemical potential. 
To keep things simple, we focus on the short-range repulsion between baryons.

Following Refs.~\cite{Vovchenko:2016rkn,Satarov:2016peb} we assume that repulsive interactions exist between all baryon-baryon and antibaryon-antibaryon pairs.
These interactions are modeled by means of the excluded-volume (EV) correction~\cite{Rischke:1991ke}. At the same time, the EV interactions between all other hadron pairs are explicitly omitted. 
It is not clearly established whether significant EV-type interactions exist between hadron pairs other than (anti)baryons~(see Ref.~\cite{Vovchenko:2016rkn} for discussion). 
We denote this setup as the EV-HRG model.
Note that this model is quite different from the usual EV prescription used in HRG model analysis: normally it is assumed that all hadrons, including mesons, have identical eigenvolume, and, thus, all hadron pairs interact repulsively at short distances~\cite{BraunMunzinger:1999qy,Cleymans:2005xv,Randrup:2009ch}.
However, a presence of a significant mesonic eigenvolume leads to notable suppression of thermodynamic functions at $\mu = 0$, which appears to be at odds with the lattice data~\cite{Andronic:2012ut,Vovchenko:2014pka}.
Note that EV corrections were recently considered also for a glueball gas in Yang-Mills theory, in the context of the corresponding lattice data~\cite{Alba:2016fku}.

The EV-HRG model consists of three independent
sub-systems: Non-interacting mesons, interacting baryons, and interacting antibaryons. 
The (anti)baryonic partial pressure $p_{_{B(\bar{B})}}^{\rm ev}$ satisfies the transcendental equation ${p_{_{B(\bar{B})}}^{\rm ev} (T, \mu_B) = p_{_{B(\bar{B})}}^{\rm id} (T, \mu_B - b \, p_{_{B\bar{B}}}^{\rm ev})}$, which can be written in the Boltzmann approximation as follows:
\eq{\label{eq:pev}
p_{_{B(\bar{B})}}^{\rm ev} (T, \mu_B) ~=~ T \, \phi_B(T) \, 
\exp\left( \frac{\mu_B - b \, p_{_{B(\bar{B})}}^{\rm ev}}{T} \right).
}
Let us denote the total densities of baryons and of antibaryons as ${n_B^{\rm ev} \equiv (\partial \, p_{_{B}}^{\rm ev} / \partial \, \mu_B)_T}$ and ${n_{\bar{B}}^{\rm ev} \equiv - (\partial \, p_{_{\bar{B}}}^{\rm ev} / \partial \, \mu_B)_T}$, respectively. By definition, the net baryon density is then $\rho_B^{\rm ev} = n_B^{\rm ev} - n_{\bar{B}}^{\rm ev}$. In the Boltzmann approximation one has the following transcendental equations for $n_B^{\rm ev}$ and $n_{\bar{B}}^{\rm ev}$~\cite{Vovchenko:2016rkn}
\eq{\label{eq:nbEV}
n_B^{\rm ev} & = (1-b\,n_B^{\rm ev}) \, \lambda_B \, \phi_B(T) \, \exp\left( - \frac{b\,n_B^{\rm ev}}{1-b\,n_B^{\rm ev}} \right), \\
n_{\bar{B}}^{\rm ev} & = (1-b\,n_{\bar{B}}^{\rm ev}) \, \lambda_B^{-1} \, \phi_B(T) \, \exp\left( - \frac{b\,n_{\bar{B}}^{\rm ev}}{1-b\,n_{\bar{B}}^{\rm ev}} \right),
}
with $\lambda_B \equiv e^{\mu_B/T}$. 
Let us assume $n_B$ and $n_{\bar{B}}$ in the following fugacity expansion form:
\eq{\label{eq:EVfug1}
\frac{n_B^{\rm ev}}{T^3} & = \frac{1}{2} \, \sum_{k=1}^{\infty} b_k^{\rm ev}(T) \, \lambda_B^k, \\
\label{eq:EVfug2}
\frac{n_{\bar{B}}^{\rm ev}}{T^3} & = \frac{1}{2} \, \sum_{k=1}^{\infty} b_k^{\rm ev}(T) \, \lambda_B^{-k}.
}
The prefactor $1/2$ is chosen such that the corresponding fugacity expansion for the net baryon density $\rho_B^{\rm ev} \equiv n_B^{\rm ev} - n_{\bar{B}}^{\rm ev}$ coincides with Eq.~\eqref{eq:chi1def}.

Putting this into Eq.~\eqref{eq:nbEV} and truncating at the fourth power of $\lambda_B$ one obtains analytic expressions for $b_k^{\rm ev}$:
\eq{\label{eq:bEV}
b_1^{\rm ev}(T) & = 2 \, \frac{\phi_B(T)}{T^3}, \\
\label{eq:bEV2}
b_2^{\rm ev}(T) & = -4  \, [b\phi_B(T)] \, \frac{\phi_B(T)}{T^3}, \\
b_3^{\rm ev}(T) & = 9 \, [b \, \phi_B(T)]^2 \, \frac{\phi_B(T)}{T^3}, \\
\label{eq:bEV4}
b_4^{\rm ev}(T) & = -\frac{64}{3} \, [b \, \phi_B(T)]^3 \, \frac{\phi_B(T)}{T^3}.
}

The first coefficient, $b_1^{\rm ev}(T)$, coincides with the ideal HRG model result. Thus, it is unaffected by the baryon-baryon EV interactions.\footnote{Note, however, that $b_1(T)$ are potentially affected by the meson-baryon EV-type interactions, which are not considered in the present work.}
Contrary to the ideal HRG model, the higher-order coefficients are non-zero. They seem to follow a generic pattern: even order coefficients are negative while odd order coefficients are positive. This sign-changing pattern was verified to be present in the EV-HRG model at least up to the 10th order. As seen from Eqs.~\eqref{eq:bEV}-\eqref{eq:bEV4}, the coefficients scale with the eigenvolume parameter as $b_k^{\rm ev} \propto b^{k-1}$. 
The ratios $b_k^{\rm ev} / \left( b_1^{\rm ev} \right)^k$ scale as $(-1)^{k+1} (b T^3)^{k-1}$, 
meaning that more and more Fourier coefficients become non-negligible as the temperature is increased.

One may also consider a more general case, where both the repulsive and also the attractive van der Waals (vdW) interactions between baryons are present. For this vdW-HRG model~\cite{Vovchenko:2016rkn} the coefficients $b_k(T)$ can also be calculated analytically. The details are given in Appendix.

\subsection{High-temperature limit of massless quarks and gluons}

Let us also mention the high-temperature limit, where the thermodynamic features of QCD are expected to resemble those of a massless ideal gas of quarks and gluons. In this Stefan-Boltzmann (SB) limit the pressure is
\eq{
\frac{p^{\rm {_{SB}}}}{T^4} = \frac{8\pi^2}{45} + \sum_{f = u,d,s} \left[\frac{7\pi^2}{60} + \frac{1}{2} \, \left( \frac{\mu_f}{T} \right)^2 + \frac{1}{4\pi^2} \, \left( \frac{\mu_f}{T} \right)^4 \right].
}
Since we only consider the case $\mu_S = \mu_Q = 0$, one has $\mu_f = \mu_B / 3$.
The net baryon susceptibility at imaginary $\mu_B$ reads
\eq{\label{eq:chi1Im}
\chi_1^B(T, i \tilde{\mu}_B) & = \left.
\frac{\partial (p/T^4)}{\partial (\mu_B / T)} \right|_{\mu_B = i \, \tilde{\mu}_B} 
\nonumber \\
& = \frac{i}{3} \, \left[ \frac{\tilde{\mu}_B}{T} - \frac{1}{9\pi^2} \left( \frac{\tilde{\mu}_B}{T} \right)^3  \right].
}
At high temperatures, Roberge-Weiss transition is expected at $\tilde{\mu}_b = \pi \, T$~\cite{Roberge:1986mm}.
Thus, the polynomial behavior given by Eq.~\eqref{eq:chi1Im} should only be considered up to this imaginary chemical potential value.

The coefficients $b_k^{\rm {_{SB}}}$ are calculated according to Eq.~\eqref{eq:bFourier}. One obtains:
\eq{\label{eq:SB}
b_k^{\rm {_{SB}}} = \frac{(-1)^{k+1}}{k} \, \frac{4 \, [3 + 4 \, (\pi k)^2]}{27 \, (\pi k)^2}.
}

The Fourier 
coefficients at very high temperatures show a sign structure: 
even coefficients are negative, odd coefficients are positive. This is exactly
the same sign structure as predicted by the EV-HRG model. On the other hand, as opposed
to the strong temperature dependence in the ratios predicted by the EV-HRG model, namely
$b_k^{\rm ev} / \left( b_1^{\rm ev} \right)^k \propto (-1)^{k+1} (b T^3)^{k-1}$,
in the free quark limit this ratio is temperature independent.

\section{Lattice method}
\label{sec:lattice}

Our lattice simulations use the tree-level Symanzik improved gauge action 
and $2+1+1$ flavours of four times stout smeared staggered quarks, with the 
smearing parameter $\rho=0.125$. The same 4stout lattice setup was also
used in~\cite{Bellwied:2015lba, Bellwied:2015rza, Gunther:2016vcp, Borsanyi:2016ksw, Alba:2017mqu}.
We use physical quark masses. 
The details of the lattice action 
can be found in~\cite{Bellwied:2015lba}.
We generate configurations with ${\textrm{Im}~\mu_B > 0}$, in the temperature 
range ${135 \leq T \leq 230}$~MeV. The geometry of our lattices is $48^3 \times 12$.
A continuum extrapolation was not attempted so far. 
We run roughly $1000$ - $2000$ configurations at each simulation point, 
separated by $10$ HMC trajectories. We measure the imaginary part $\chi^B_1$
on the lattices, and carry out a discrete Fourier transform to obtain the
observables $b_1$, $b_2$, $b_3$ and $b_4$. 
The errors on the lattice data points are purely statistical, calculated from 48 jackknife samples.

The crucial observation is that the Fourier coefficients at imaginary 
chemical potential correspond to partial pressures coming from different 
sectors of the Hilbert space. 
These can also be identified with the fugacity or relativistic virial expansion coefficients for real chemical potential.
From a phenomenological point of view, this 
makes the Fourier coefficients particularly sensitive to the details of
hadronic models. This was pointed out in ~\cite{Alba:2017mqu}, where
the different strangeness sectors of the theory were separated, and
later used to constrain the hadronic spectrum in the context of the ideal HRG model.
Note that the fugacity expansion of the logarithm of the partition function, $\log \textrm{Z}$,
employed in the present work, is quite different from the fugacity expansion of the fermion determinant, which corresponds to the fugacity expansion of $\textrm{Z}$ and which had also been used in some lattice studies~\cite{Danzer:2012vw,Gattringer:2014hra}.
Strong finite volume scaling effects in the fugacity expansion of $\log \textrm{Z}$ are not expected, in contrast to the fugacity expansion of $\textrm{Z}$.

Other studies~\cite{Huovinen:2017ogf} exploit the connection of the virial coefficients to the fluctuations of conserved charges at $\mu=0$. 
E.g., if one neglects the third and higher order coefficients in the expansion~\eqref{eq:pseries}, i.e. $a_3=a_4=\dots=0$, then the difference $\chi^B_4-\chi^B_2$ of the fourth and second order baryon susceptibilities is simply proportional to the second coefficient $a_2$.
The validity of the truncation to only the second coefficient breaks down as the temperature is increased, and such a method no longer works correctly.
In this work we use a different approach, and calculate the expansion coefficients directly, by exploiting the fact that they become Fourier coefficients at an imaginary chemical potential. 
This allows us to consider the higher order coefficients as well, apart from $b_2$. 
Moreover, in~\cite{Alba:2017mqu} we show by explicit lattice calculations of the strangeness sectors in the confined phase, that in the cases where the truncation of the virial expansion is warranted, and the two methods should agree, our method produces smaller statistical errors for the same computational cost.
We note that coefficients $b_k$ were considered in the lattice studies before~(see e.g. Refs.~\cite{DElia:2007bkz,Takahashi:2014rta,Bornyakov:2016wld,Boyda:2017dyo}), where they were estimated by fitting the lattice data with the truncated fugacity expansion.

\section{Results and discussion}
\label{sec:results}

\subsection{Hadronic description}

\begin{figure*}[t]
  \centering 
  \includegraphics[width=.49\textwidth]{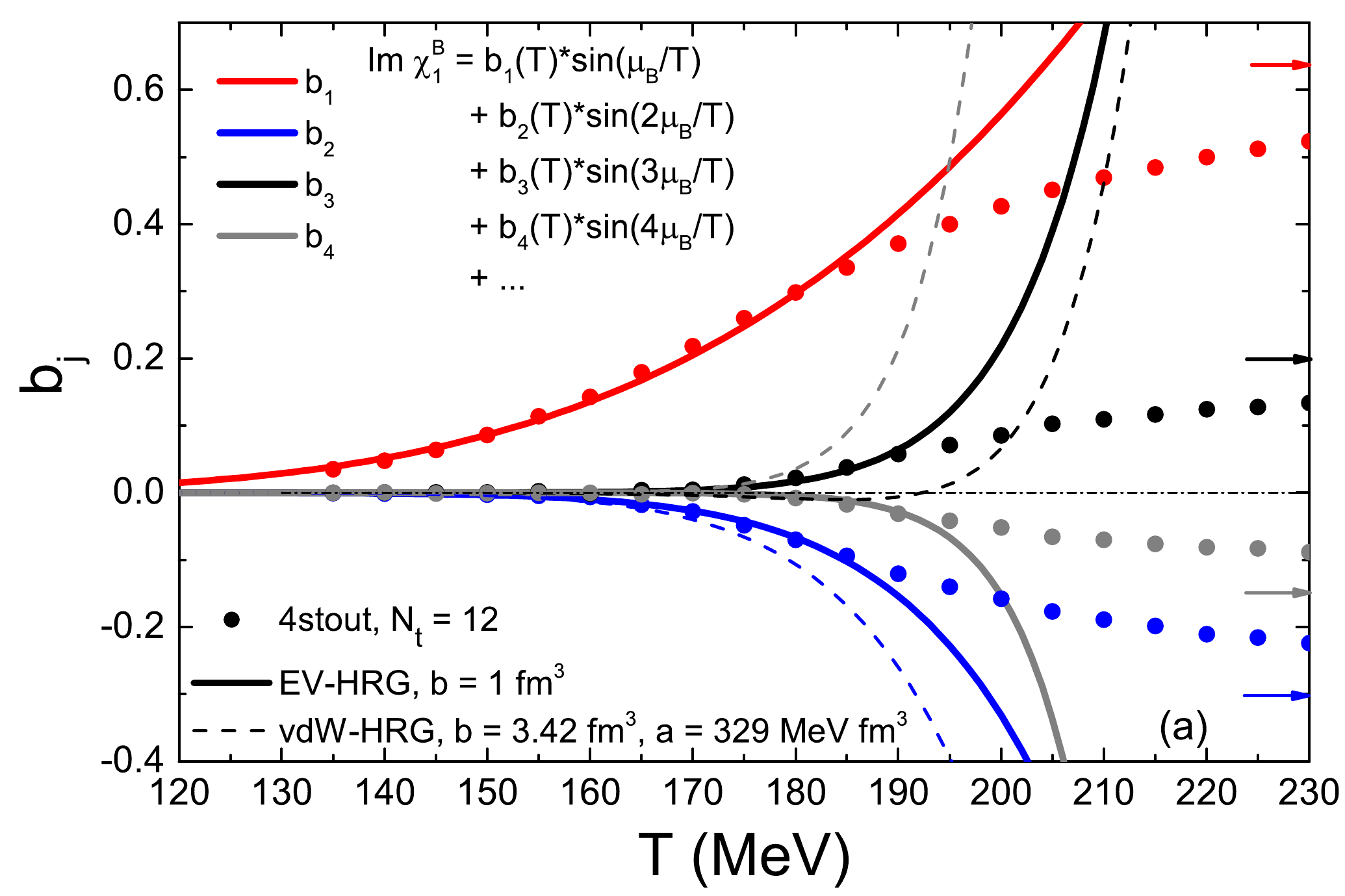}
  \includegraphics[width=.49\textwidth]{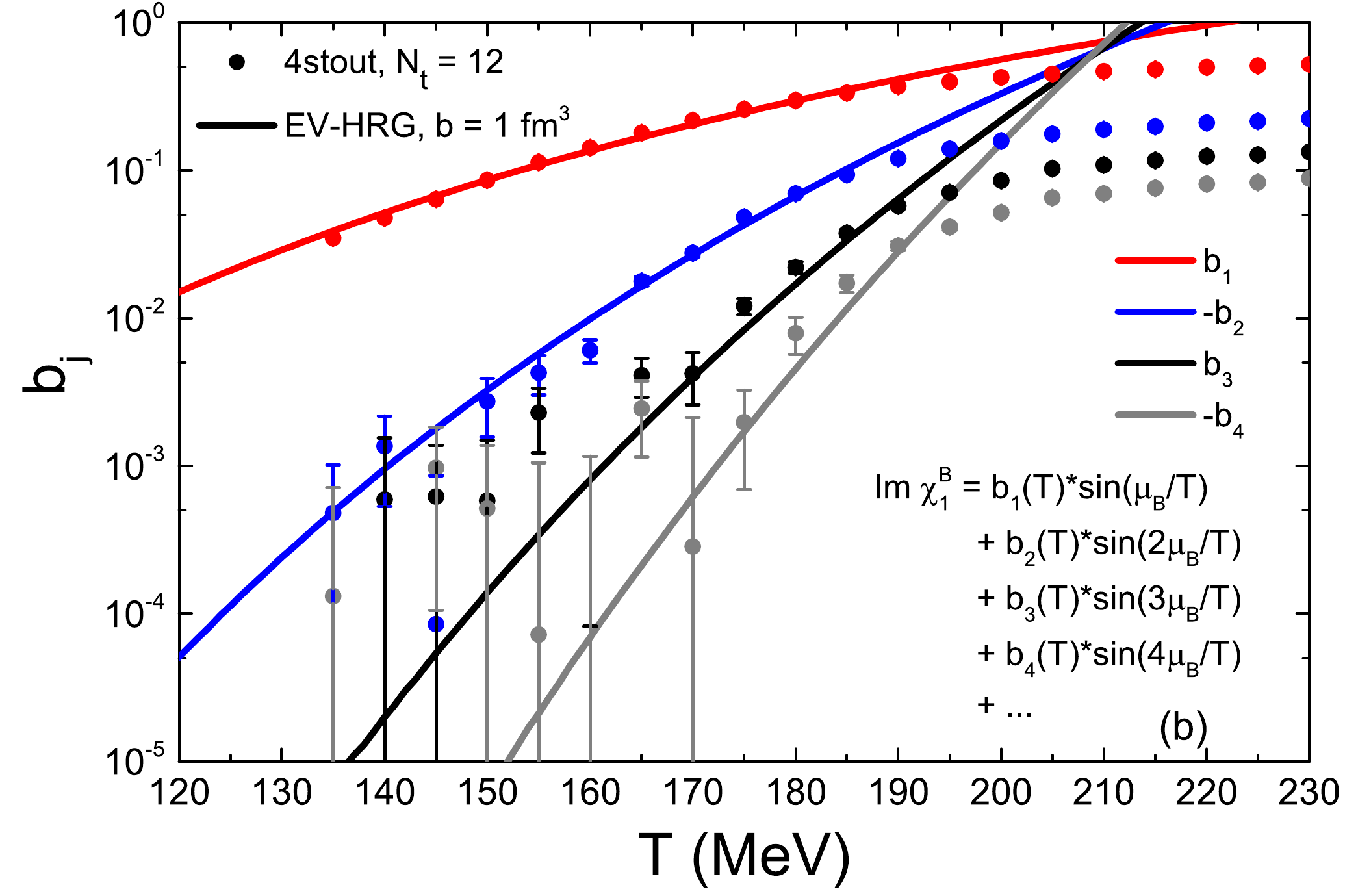}
  \caption{ The temperature dependence of the first four Fourier coefficients $b_k$~\eqref{eq:chi1imag}, calculated on the lattice with the 4stout, $N_t = 12$ setup (symbols), and within the EV-HRG model with baryonic eigenvolume parameter $b = 1$~fm$^3$ (solid lines). This dependence is shown on (a) the linear and (b) the logarithmic scales. The dashed lines in (a) show the calculations within the vdW-HRG model, with van der Waals parameters $a$ and $b$ fixed by the properties of the nuclear ground state~\cite{Vovchenko:2016rkn}.
  The arrows in (a) correspond to the Stefan-Boltzmann limit~\eqref{eq:SB} of the massless gas of quarks and gluons.} 
  \label{fig:EVfit}
\end{figure*}

Figure~\ref{fig:EVfit} depicts the temperature dependence of the first four Fourier coefficients $b_k$~\eqref{eq:chi1imag}, calculated on the lattice with the 4stout, $N_t = 12$ setup (symbols), and within the EV-HRG model with baryonic eigenvolume parameter $b = 1$~fm$^3$ (solid lines). This dependence is shown (a) on the linear scale, and (b) on the logarithmic scale.

As mentioned above, the $b_1$ coefficient is not affected by the baryon-baryon interactions. 
Its behavior in the EV-HRG model is the same as in the ideal HRG model, and it is determined solely by the input particle list and, less so, by the modeling of the finite resonance widths.
The HRG model with the PDG-based hadron list, employed in the present work, 
stays rather close to the lattice data for $b_1$ up to $T \simeq 185$~MeV,
but does not reproduce the inflection, and therefore does not describe the temperature derivative of the $b_1(T)$ curve
well from $T \simeq 175$~MeV.

Lattice calculations predict non-zero values for the higher-order coefficients. For instance, the second coefficient $b_2$ is negative in the considered temperature range. As seen from Fig.~\ref{fig:EVfit}(a), this coefficient starts to notably deviate from zero at about $T \simeq 160$~MeV. This deviation signals the end of the applicability range of the ideal HRG model, which predicts $b_2 \equiv 0$ at all temperatures. 

The negative sign is expected in the case where the second Fourier coefficient is dominated by the elastic
two-to-two baryon-baryon scattering with a repulsive interaction. In this case the second Fourier coefficient is given by the Beth-Uhlenbeck formula~\cite{Dashen:1969ep,Beth:1937zz},
and its sign is therefore given by the sign of the energy derivative of the scattering phase shift, which is negative in the case of a repulsive interaction.

The third and fourth order coefficients, as calculated on the lattice, start to notably deviate from zero at successively higher temperatures. Lattice calculations show a peculiar alternating sign hierarchy: odd order coefficients, $b_1$ and $b_3$, are positive while the even order coefficients, $b_2$ and $b_4$, are 
negative\footnote{This proliferation of Fourier coefficients at high temperature can also be regarded
as a signal for the Roberge-Weiss transition~\cite{Roberge:1986mm}.}.
We note that indications for such behavior of the first four coefficients were already seen in lattice simulations before~\cite{DElia:2007bkz,Takahashi:2014rta,Bornyakov:2016wld,Boyda:2017dyo}, and, in particular, the alternating sign structure of the first four coefficients was obtained in Ref.~\cite{DElia:2007bkz}.
Interestingly, this structure is also predicted by the EV-HRG model with repulsive baryonic interactions, as seen in Eqs.~\eqref{eq:bEV}-\eqref{eq:bEV4}.
In fact, the EV-HRG model with appropriately chosen baryonic eigenvolume parameter describes the lattice data fairly well: as seen in Fig.~\ref{fig:EVfit}, all four coefficients calculated in the EV-HRG model with $b = 1$~fm$^3$ are in good agreement with the lattice data at temperatures $T \lesssim 185$~MeV.
Thus, such a choice of the $b$ value includes many of the non-perturbative corrections, which are otherwise very complicated.
The lattice results for $b_k$ do contain the inflection points in the temperature dependence, which are not predicted by the EV-HRG model. 
All four coefficients, as calculated on the lattice, appear to converge slowly towards the corresponding Stefan-Boltzmann limiting values, which are given by Eq.~\eqref{eq:SB}.

For completeness, we also depict the results obtained within the vdW-HRG model~\cite{Vovchenko:2016rkn}, with vdW parameters $a$ and $b$ extracted from the nuclear ground state properties.
Unlike EV-HRG model, this model describes correctly the basic binding properties of nuclear matter at low temperatures and high baryochemical potentials, and it has no free parameters which can be adjusted to fit lattice data. The vdW-HRG model gives a fair description of $b_2$ at lower temperatures, but misses the $b_3$ and $b_4$.
It appears that nuclear matter based values of vdW parameters, namely $a = 329$~MeV~fm$^3$ and $b = 3.42$~fm$^3$, are overestimated when applied to the description of the lattice data at $T = 130-190$~MeV. The EV-HRG model with a smaller $b = 1$~fm$^3$ does a much better job in describing the lattice data. It would be interesting to reconcile both models, and obtain a simultaneous description of the lattice data and of the nuclear matter properties.

\begin{figure}[t]
  \centering 
  \includegraphics[width=.49\textwidth]{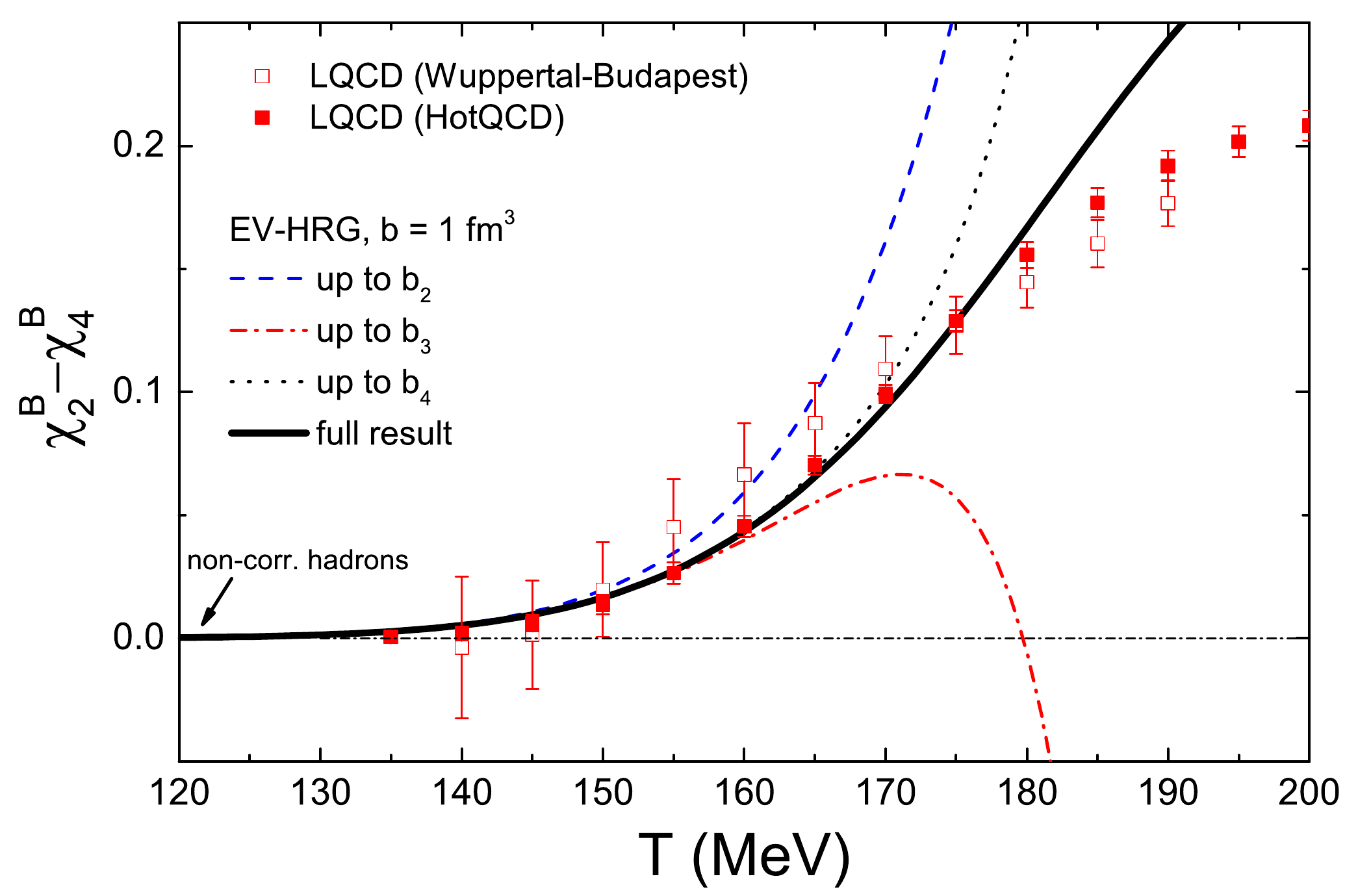}
  \caption{The temperature dependence of $\chi_2^B - \chi_4^B$, calculated within the EV-HRG model with baryonic eigenvolume parameter $b = 1$~fm$^3$~(solid line).
Other lines depict the EV-HRG model calculations using the fugacity expansion~\eqref{eq:chi2chi4def}, truncated at the second~(dashed blue line), third~(dash-dotted red line), and fourth~(dotted black line) orders.
 Lattice QCD data from Refs.~\cite{Bellwied:2015lba} and~\cite{Bazavov:2017dus,Bazavov:2017tot} are depicted, respectively, by open and full symbols.
  } 
  \label{fig:chi2chi4}
\end{figure}

The Fourier expansion coefficients $b_k(T)$ can be contrasted with the net baryon number susceptibilities $\chi_k^B(T)$ at zero baryochemical potential, which are defined as follows:
\eq{\label{eq:chikdef}
\chi_k^B(T) \equiv \left. \frac{\partial^k (p/T^4)}{\partial (\mu_B / T)^k} \right|_{\mu_B = 0}.
}
$\chi_k^B$ are proportional to the coefficients of the Taylor expansion of the QCD pressure with respect to $\mu_B$, they correspond to the cumulants of the baryon number distribution at a given temperature and, therefore, they are more directly connected to the observables which are measured in heavy-ion collision experiments.
It is particularly instructive to consider the difference $\chi_2^B - \chi_4^B$.
The fugacity expansion~\eqref{eq:chi1def} for this quantity reads
\eq{\label{eq:chi2chi4def}
\chi_2^B - \chi_4^B = -\sum_{k=2}^\infty \, k \, (k^2 - 1) \, b_k(T)~.
}
In the ideal HRG this quantity is strictly zero. 
This is no longer the case when baryonic interactions are included. 
When the effects of baryon-baryon interactions are small, the third and higher order coefficients in Eq.~\eqref{eq:chi2chi4def} can be neglected. 
In this case $\chi_2^B - \chi_4^B$ is directly proportional to $b_2$, which is in turn proportional to the second virial coefficient of baryon-baryon interactions.
This fact was exploited in Ref.~\cite{Huovinen:2017ogf}.
When the total density of baryons is high, however, higher order terms of the expansion~\eqref{eq:chi2chi4def} have to be considered as well.

We calculate the temperature dependence of the difference $\chi_2^B - \chi_4^B$ using the EV-HRG model with $b = 1$~fm$^3$.
To study the breakdown of the truncated fugacity expansion at high temperatures, we consider the expansion~\eqref{eq:chi2chi4def}, which is truncated at the second, third, or fourth order. 
It is calculated using Eqs.~\eqref{eq:bEV2}-\eqref{eq:bEV4} for the $b_k$ coefficients in the EV-HRG model.
These calculations are compared with the full, untruncated result, obtained by directly solving the transcendental equation~\eqref{eq:pev} for the pressure and using the definition~\eqref{eq:chikdef} for $\chi_k^B$.
Lattice QCD results from Refs.~\cite{Bellwied:2015lba} and~\cite{Bazavov:2017dus,Bazavov:2017tot} are also shown.

Results exhibited in Fig.~\ref{fig:chi2chi4} demonstrate the validity range for the different orders of truncation used for calculating $\chi_2^B - \chi_4^B$.
The second order works well up to $T \simeq 150$~MeV, the third order is applicable up to $T \simeq 160$~MeV, and the fourth order reproduces the full result until $T \simeq 170$~MeV.
We note that the validity range of a particular truncation scheme can be different if used for a different observable.
Our above conclusions apply specifically to $\chi_2^B - \chi_4^B$.

As seen from Fig.~\ref{fig:chi2chi4}, the full EV-HRG model reproduces the lattice data quite well up to $T \simeq 175$~MeV.
The non-zero values of the $\chi_2^B - \chi_4^B$ difference were suggested as a possible indicator of deconfinement in~\cite{Bazavov:2013dta}, our analysis suggests an alternative possibility in terms of repulsive baryonic interactions.
The model predictions are no longer consistent with the lattice data at $T > 185$~MeV. For example, it was checked that $\chi_4^B$ becomes negative at $T \simeq 187$~MeV, a behavior not seen in lattice simulations.

The success of the EV-HRG model in describing the Fourier coefficients and the baryon number susceptibilities does not automatically mean that such a model describes all other QCD observables, for instance the correlations and fluctuations involving the electric charge and strangeness, in the same temperature range. 
These observables are sensitive to the baryon-baryon interactions  as well~\cite{Vovchenko:2016rkn}.
At the same time, they are also sensitive to interactions involving mesons, as these carry both electric charge and strangeness,
and to the strangeness-dependent baryonic interactions~\cite{Vovchenko:2017zpj,Huovinen:2017ogf}.
These extensions are beyond the scope of the present paper.

\subsection{Parameters extracted from lattice}

\begin{figure}[t]
  \centering 
  \includegraphics[width=.49\textwidth]{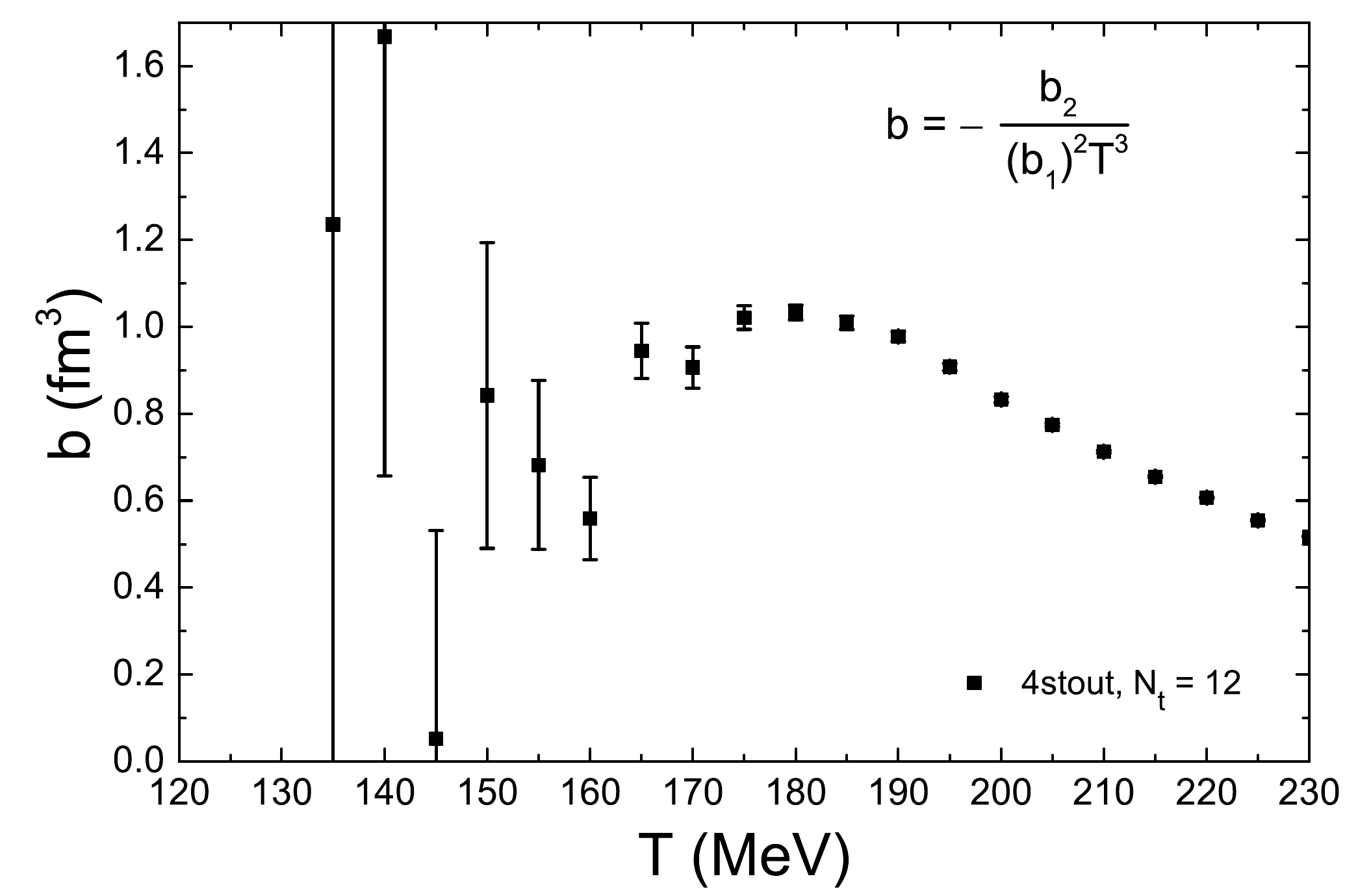}
  \caption{The temperature dependence of the "eigenvolume parameter" $b(T)$, as estimated from the 
      lattice according to Eq.~\eqref{eq:bTdep}.
  } 
  \label{fig:EVbTdep}
\end{figure}

We consider a modification of the EV-HRG model, where the first two 
Fourier coefficients, i.e. the partial pressures from the $|B|=1$ and
$|B|=2$ sectors, are treated as temperature dependent free parameters,
and are fitted to the lattice data. This corresponds to calculating
the functions $\phi_B(T)$ and $b(T)$, defined by:
\eq{\label{eq:phibTdep}
    \phi_B(T) = \frac{b_1(T)}{2} T^3,
}
and
\eq{\label{eq:bTdep}
    b(T) = - \frac{b_2 (T)}{[b_1(T)]^2} \frac{1}{T^3},
}
where $b_1(T)$ and $b_2(T)$ are taken from lattice simulations.
These relations follow from Eqs.~\eqref{eq:bEV} and \eqref{eq:bEV2}.
The lattice-extracted $b(T)$ is plotted in Fig.~\ref{fig:EVbTdep}.

The values of $b(T)$ are fairly consistent with 1~fm$^3$ at $T < 190$~MeV.
It is interesting that $b(T)$ shows plateau slightly above the pseudocritical temperature.
$b(T)$ monotonously decreases at high temperatures, in the regions where one does not expect to find hadrons in their normal form.
In fact, to reproduce the asymptotic expectation of the $b_k/(b_1)^k$ ratios being independent of temperature, the
parameter $b$ has to 
scale as
$b \propto 1/T^3$ at high temperatures.

Of course, the estimate plotted in Fig.~\ref{fig:EVbTdep} is only a model-dependent interpretation of the lattice data, which should be treated with care.
This scenario corresponds to a hadronic description with eigenvolume interactions for baryon-baryon and for antibaryon-antibaryon pairs, while all other hadron pairs are considered to be non-interacting.
In general, even for the purely hadronic description, the $b(T)$ values extracted from the lattice reflect the net contribution to the 2nd virial coefficient of both the repulsive and the attractive baryonic interactions.
This contribution is averaged over all baryon-baryon pairs. 
Thus, it cannot distinguish possible differences in virial coefficients for different baryon pairs, for instance involving the strange baryons~\cite{Vovchenko:2017zpj}.
If the attractive interactions are non-negligible, then $b(T)$ cannot be attributed exclusively to the baryonic eigenvolume.
Since the nucleon-nucleon interaction is attractive at the intermediate range, it is predicted that $b(T)$ should become negative at sufficiently small temperatures, where the hadron gas is dilute and where the average distance between baryons becomes larger.

\begin{figure}[t]
  \centering 
  \includegraphics[width=.49\textwidth]{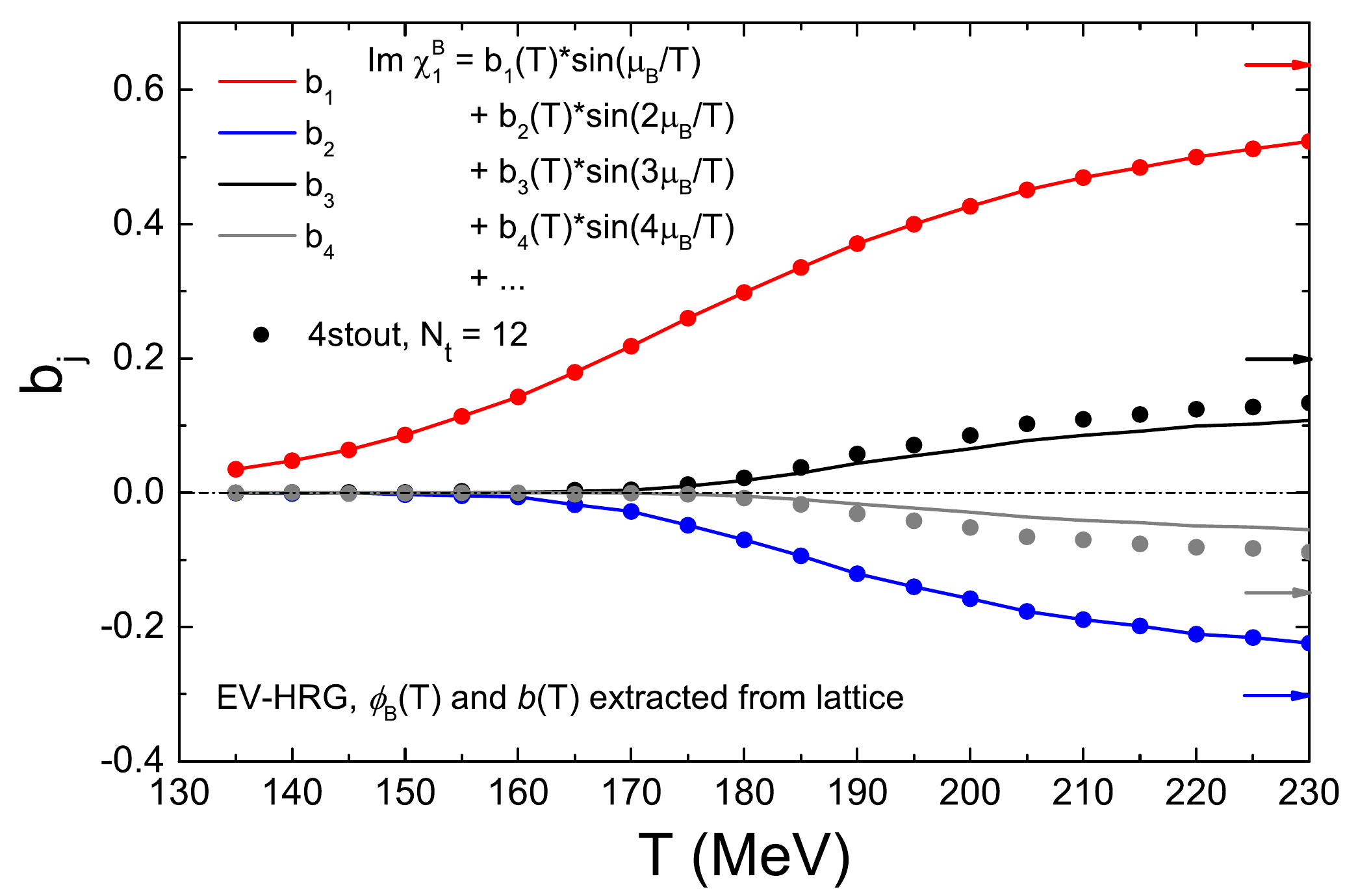}
  \caption{Same as Fig.~\ref{fig:EVfit}, but calculated for the modified EV-HRG model, where the first two Fourier coefficients are not taken from standard EV-HRG, but are tuned to exactly reproduce the lattice data by construction, using Eqs.~\eqref{eq:phibTdep} and \eqref{eq:bTdep}, and the higher coefficients are calculated from these $b_1$ and $b_2$ using the formulas given by the EV-HRG model.
} 
\label{fig:EVlatt}
\end{figure}

At high temperatures, $T > 185$~MeV, the lattice data for the $b_1$ coefficient cannot be described by the standard baryonic spectrum in HRG, as seen from Fig.~\ref{fig:EVfit}.
The function $\phi_B$, as extracted from the lattice with Eq.~\eqref{eq:phibTdep}, no longer reflects baryons in their normal vacuum form.
Nevertheless, it is interesting that the EV-HRG model with the lattice-extracted $\phi_B(T)$ and $b(T)$ gives a reasonable description of the $b_3$ and $b_4$ coefficients even at temperatures $T > 200$~MeV~(Fig.~\ref{fig:EVlatt}). 
This is quite notable since the the $b_3$ and $b_4$ coefficients are not used to extract $\phi_B(T)$ and $b(T)$ from the lattice.
One can expect a similar picture for the higher order coefficients, which define the properties of the more and more dense baryon medium.
The result suggests that the EV-HRG model has a certain predictive power,
particularly regarding the baryon-rich region of the phase diagram, which is presently unaccessible by the lattice simulations.
These questions will be explored in the future studies.

\section{Summary}
\label{sec:summary}

We presented the lattice QCD observables at an imaginary baryochemical potential, and analyzed them in the framework of a hadron resonance gas model with repulsive interactions between baryons. More specifically, the temperature dependent Fourier coefficients of the Fourier series expansion of the net baryon density at imaginary $\mu_B$ were considered. The ideal hadron resonance gas model predicts zero values for the 2nd and higher-order coefficients.
Thus, significant deviations from zero of the higher-order coefficients signal the end of the applicability of the ideal HRG model.
Lattice calculations predict that the onset of this behavior takes place at about $T = 160$~MeV. They also predict an alternating sign structure for the coefficients: the odd order coefficients, $b_1$ and $b_3$ are positive, while the even order ones, $b_2$ and $b_4$, are negative.

Remarkably, the behavior of the first four Fourier coefficients at $T \lesssim 185$~MeV appears to be well described by the HRG model with the excluded-volume interactions between baryons, characterized by a single eigenvolume parameter $b \simeq 1$~fm$^3$. We do note that some finer structures, such as the temperature derivatives of the coefficients, 
or the difference $\chi_2^B - \chi_4^B$ of baryon number susceptibilities at $\mu_B = 0$,
are reproduced by this simple model only up to a lower temperature of about 175~MeV. 
The EV-HRG model also predicts the alternating sign structure analytically. At the same time, the van der Waals HRG model, with vdW parameters $a$ and $b$ fixed by the properties of the nuclear ground state, does a worse job in describing the Fourier coefficients. It will be interesting to reconcile these two approaches in order to obtain a unified model for the hadronic equation of state. This model would describe both, the nuclear matter properties at low temperatures and high baryon densities, and the lattice QCD data at high temperatures.
A proper hadronic baseline is crucially important for the ongoing experimental effort in determining the properties of QCD from the heavy-ion collisions experiments at different collision energies.

The present study elucidates the potential of the lattice QCD observables at imaginary chemical potentials to shed light on the properties of QCD, particularly regarding the hadronic interactions in the confined phase. Such analysis should also be performed for other imaginary $\mu$ observables, e.g. involving the electric charge and strangeness, as well as for the more accurate, and continuum extrapolated lattice data, which will be available in the future\footnote{
In the final stages of preparation of this paper, Ref.~\cite{Huovinen:2017ogf} appeared on arXiv. That paper covers a similar topic, but with a different method. 
Instead of imaginary $\mu$, the $\mu=0$ simulations are used there,
and, unlike present work, only the first $b_2$ correction to the ideal HRG model is considered in~\cite{Huovinen:2017ogf}.
}.

%TC:ignore 

\begin{acknowledgments}

%\section*{Acknowledgements}
%\emph{Acknowledgments.} 
We are grateful to Szabolcs Bors\'anyi for stimulating discussions and for his help with the lattice data.
This work was supported by HIC for FAIR within the LOEWE program of the State of Hesse.
V.V. acknowledges the support from HGS-HIRe for FAIR.
H.St. acknowledges the support through the Judah M. Eisenberg Laureatus Chair at Goethe University.
This project was funded by the DFG grant SFB/TR55.
This research used resources of the Argonne
Leadership Computing Facility, which is a DOE Office
of Science User Facility supported under Contract DEAC02-06CH11357.
The authors gratefully acknowledge
the Gauss Centre for Supercomputing (GCS) for providing
computing time for a GCS Large-Scale Project on
the GCS share of the supercomputer JUQUEEN~\cite{juqueen} at
J\"ulich Supercomputing Centre (JSC), and at HazelHen
supercomputer at HLRS, Stuttgart.

\end{acknowledgments}

\begin{appendix}

\section*{Appendix}
%\appendix

This appendix presents the calculation of the Fourier coefficients $b_k(T)$ in the Fourier expansion of the net baryon susceptibility $\chi_1^B$ at an imaginary baryochemical potential~\eqref{eq:chi1imag} for the vdW-HRG model~\cite{Vovchenko:2016rkn}. In the vdW-HRG model, the attractive and repulsive baryonic interactions are described by the van der Waals equation, with common $a$ and $b$ parameters for all baryons. For $a=0$ this model reduces to the EV-HRG model in Sec.~\ref{sec:EVHRG}.

Following results of Ref.~~\cite{Vovchenko:2016rkn}, in the Boltzmann approximation one has the following transcendental equations for $n_B^{\rm vdw}$ and $n_{\bar{B}}^{\rm vdw}$
\eq{\label{eq:nbVDW}
n_B^{\rm vdw} & = (1-b\,n_B^{\rm vdw}) \, \lambda_B \, \phi_B(T) \, \exp\left( - \frac{b\,n_B^{\rm vdw}}{1-b\,n_B^{\rm vdw}} \right) \nonumber \\
& \quad \times \exp\left( \frac{2\,a\,n_B^{\rm vdw}}{T} \right), \\
n_{\bar{B}}^{\rm vdw} & = (1-b\,n_{\bar{B}}^{\rm vdw}) \, \lambda_B^{-1} \, \phi_B(T) \, \exp\left( - \frac{b\,n_{\bar{B}}^{\rm vdw}}{1-b\,n_{\bar{B}}^{\rm vdw}} \right) \nonumber \\
& \quad  \times \exp\left( \frac{2\,a\,n_{\bar{B}}^{\rm vdw}}{T} \right).
}

The calculation of the coefficients $b_k^{\rm vdw}$ proceeds in essentially the same way as it was done for the EV-HRG model.
One assumes the fugacity expansions for $n_{B(\bar{B})}^{\rm vdw}$ in the form \eqref{eq:EVfug1}-\eqref{eq:EVfug2}, and calculates the $b_k^{\rm vdw}$ by plugging in the fugacity expansion into Eq.~\eqref{eq:nbVDW}. The result is
\eq{\label{eq:bkVDW}
b_1^{\rm vdw}(T) & = 2 \, \frac{\phi_B(T)}{T^3}, \\
b_2^{\rm vdw}(T) & = -4  \, \left(b - \frac{a}{T}\right) \, \phi_B(T) \, \frac{\phi_B(T)}{T^3}, \\
b_3^{\rm vdw}(T) & = 9 \, \left(b^2 - \frac{8}{3} \, \frac{a\,b}{T}  + \frac{4}{3} \, \frac{a^2}{T^2} \right)\, [\phi_B(T)]^2 \, \frac{\phi_B(T)}{T^3}, \\
b_4^{\rm vdw}(T) & = -\frac{64}{3} \, \left(b^3 - \frac{39}{8} \, \frac{a\,b^2}{T}  + 6 \, \frac{a^2 \, b}{T^2} - 2 \frac{a^3}{T^3} \right) \nonumber \\ 
& \quad \times [\phi_B(T)]^3 \, \frac{\phi_B(T)}{T^3}.
}

\end{appendix}

%TC:endignore 


\begin{thebibliography}{999}

%\cite{Borsanyi:2013bia}
\bibitem{Borsanyi:2013bia} 
  S.~Borsanyi, Z.~Fodor, C.~Hoelbling, S.~D.~Katz, S.~Krieg and K.~K.~Szabo,
  %``Full result for the QCD equation of state with 2+1 flavors,''
  Phys.\ Lett.\ B {\bf 730}, 99 (2014)
  %doi:10.1016/j.physletb.2014.01.007
  [arXiv:1309.5258 [hep-lat]].
  %%CITATION = doi:10.1016/j.physletb.2014.01.007;%%
  %295 citations counted in INSPIRE as of 07 Jul 2017


%\cite{Bazavov:2014pvz}
\bibitem{Bazavov:2014pvz} 
  A.~Bazavov {\it et al.} [HotQCD Collaboration],
  %``Equation of state in ( 2+1 )-flavor QCD,''
  Phys.\ Rev.\ D {\bf 90}, 094503 (2014)
  %doi:10.1103/PhysRevD.90.094503
  [arXiv:1407.6387 [hep-lat]].
  %%CITATION = doi:10.1103/PhysRevD.90.094503;%%
  %324 citations counted in INSPIRE as of 07 Jul 2017


%\cite{Aoki:2006we}
\bibitem{Aoki:2006we} 
  Y.~Aoki, G.~Endrodi, Z.~Fodor, S.~D.~Katz and K.~K.~Szabo,
  %``The Order of the quantum chromodynamics transition predicted by the standard model of particle physics,''
  Nature {\bf 443}, 675 (2006)
  %doi:10.1038/nature05120
  [hep-lat/0611014].
  %%CITATION = doi:10.1038/nature05120;%%
  %961 citations counted in INSPIRE as of 07 Jul 2017

%\cite{Borsanyi:2010bp}
\bibitem{Borsanyi:2010bp} 
  S.~Borsanyi {\it et al.} [Wuppertal-Budapest Collaboration],
  %``Is there still any T_c mystery in lattice QCD? Results with physical masses in the continuum limit III,''
  JHEP {\bf 1009}, 073 (2010)
  %doi:10.1007/JHEP09(2010)073
  [arXiv:1005.3508 [hep-lat]].
  %%CITATION = doi:10.1007/JHEP09(2010)073;%%
  %654 citations counted in INSPIRE as of 13 Sep 2017

%\cite{Bazavov:2011nk}
\bibitem{Bazavov:2011nk} 
  A.~Bazavov {\it et al.},
  %``The chiral and deconfinement aspects of the QCD transition,''
  Phys.\ Rev.\ D {\bf 85}, 054503 (2012)
  %doi:10.1103/PhysRevD.85.054503
  [arXiv:1111.1710 [hep-lat]].
  %%CITATION = doi:10.1103/PhysRevD.85.054503;%%
  %636 citations counted in INSPIRE as of 13 Sep 2017


%\cite{Bellwied:2013cta}
\bibitem{Bellwied:2013cta} 
  R.~Bellwied, S.~Borsanyi, Z.~Fodor, S.~D.~Katz and C.~Ratti,
  %``Is there a flavor hierarchy in the deconfinement transition of QCD?,''
  Phys.\ Rev.\ Lett.\  {\bf 111}, 202302 (2013)
  %doi:10.1103/PhysRevLett.111.202302
  [arXiv:1305.6297 [hep-lat]].
  %%CITATION = doi:10.1103/PhysRevLett.111.202302;%%
  %86 citations counted in INSPIRE as of 07 Jul 2017

%\cite{Borsanyi:2011sw}
\bibitem{Borsanyi:2011sw} 
  S.~Borsanyi, Z.~Fodor, S.~D.~Katz, S.~Krieg, C.~Ratti and K.~Szabo,
  %``Fluctuations of conserved charges at finite temperature from lattice QCD,''
  JHEP {\bf 1201}, 138 (2012)
  %doi:10.1007/JHEP01(2012)138
  [arXiv:1112.4416 [hep-lat]].
  %%CITATION = doi:10.1007/JHEP01(2012)138;%%
  %201 citations counted in INSPIRE as of 07 Jul 2017


%\cite{Bazavov:2012jq}
\bibitem{Bazavov:2012jq} 
  A.~Bazavov {\it et al.} [HotQCD Collaboration],
  %``Fluctuations and Correlations of net baryon number, electric charge, and strangeness: A comparison of lattice QCD results with the hadron resonance gas model,''
  Phys.\ Rev.\ D {\bf 86}, 034509 (2012)
  %doi:10.1103/PhysRevD.86.034509
  [arXiv:1203.0784 [hep-lat]].
  %%CITATION = doi:10.1103/PhysRevD.86.034509;%%
  %184 citations counted in INSPIRE as of 07 Jul 2017


%\cite{Bellwied:2015lba}
\bibitem{Bellwied:2015lba} 
  R.~Bellwied, S.~Borsanyi, Z.~Fodor, S.~D.~Katz, A.~Pasztor, C.~Ratti and K.~K.~Szabo,
  %``Fluctuations and correlations in high temperature QCD,''
  Phys.\ Rev.\ D {\bf 92}, 114505 (2015)
  %doi:10.1103/PhysRevD.92.114505
  [arXiv:1507.04627 [hep-lat]].
  %%CITATION = doi:10.1103/PhysRevD.92.114505;%%
  %50 citations counted in INSPIRE as of 07 Jul 2017


%\cite{Vovchenko:2016rkn}
\bibitem{Vovchenko:2016rkn} 
  V.~Vovchenko, M.~I.~Gorenstein and H.~Stoecker,
  %``van der Waals Interactions in Hadron Resonance Gas: From Nuclear Matter to Lattice QCD,''
  Phys.\ Rev.\ Lett.\  {\bf 118},  182301 (2017)
  %doi:10.1103/PhysRevLett.118.182301
  [arXiv:1609.03975 [hep-ph]].
  %%CITATION = doi:10.1103/PhysRevLett.118.182301;%%
  %6 citations counted in INSPIRE as of 07 Jul 2017

%\cite{Vovchenko:2017cbu}
\bibitem{Vovchenko:2017cbu} 
  V.~Vovchenko,
  %``Equations of state for real gases on the nuclear scale,''
  Phys.\ Rev.\ C {\bf 96}, 015206 (2017)
  %doi:10.1103/PhysRevC.96.015206
  [arXiv:1701.06524 [nucl-th]].
  %%CITATION = doi:10.1103/PhysRevC.96.015206;%%
  %3 citations counted in INSPIRE as of 05 Aug 2017

%\cite{Alba:2014eba}
\bibitem{Alba:2014eba} 
  P.~Alba, W.~Alberico, R.~Bellwied, M.~Bluhm, V.~Mantovani Sarti, M.~Nahrgang and C.~Ratti,
  %``Freeze-out conditions from net-proton and net-charge fluctuations at RHIC,''
  Phys.\ Lett.\ B {\bf 738}, 305 (2014)
  %doi:10.1016/j.physletb.2014.09.052
  [arXiv:1403.4903 [hep-ph]].
  %%CITATION = doi:10.1016/j.physletb.2014.09.052;%%
  %89 citations counted in INSPIRE as of 27 Jul 2017

%\cite{Alba:2015iva}
\bibitem{Alba:2015iva} 
  P.~Alba, R.~Bellwied, M.~Bluhm, V.~Mantovani Sarti, M.~Nahrgang and C.~Ratti,
  %``Sensitivity of multiplicity fluctuations to freeze-out conditions in heavy ion collisions,''
  Phys.\ Rev.\ C {\bf 92}, 064910 (2015)
  %doi:10.1103/PhysRevC.92.064910
  [arXiv:1504.03262 [hep-ph]].
  %%CITATION = doi:10.1103/PhysRevC.92.064910;%%
  %14 citations counted in INSPIRE as of 27 Jul 2017

%\cite{Barbour:1997ej}
\bibitem{Barbour:1997ej} 
  I.~M.~Barbour, S.~E.~Morrison, E.~G.~Klepfish, J.~B.~Kogut and M.~P.~Lombardo,
  %``Results on finite density QCD,''
  Nucl.\ Phys.\ Proc.\ Suppl.\  {\bf 60A}, 220 (1998)
  %doi:10.1016/S0920-5632(97)00484-2
  [hep-lat/9705042].
  %%CITATION = doi:10.1016/S0920-5632(97)00484-2;%%
  %162 citations counted in INSPIRE as of 07 Jul 2017


%\cite{Fodor:2001au}
\bibitem{Fodor:2001au} 
  Z.~Fodor and S.~D.~Katz,
  %``A New method to study lattice QCD at finite temperature and chemical potential,''
  Phys.\ Lett.\ B {\bf 534}, 87 (2002)
  %doi:10.1016/S0370-2693(02)01583-6
  [hep-lat/0104001].
  %%CITATION = doi:10.1016/S0370-2693(02)01583-6;%%
  %455 citations counted in INSPIRE as of 07 Jul 2017


%\cite{Fodor:2001pe}
\bibitem{Fodor:2001pe} 
  Z.~Fodor and S.~D.~Katz,
  %``Lattice determination of the critical point of QCD at finite T and mu,''
  JHEP {\bf 0203}, 014 (2002)
  %doi:10.1088/1126-6708/2002/03/014
  [hep-lat/0106002].
  %%CITATION = doi:10.1088/1126-6708/2002/03/014;%%
  %744 citations counted in INSPIRE as of 07 Jul 2017


%\cite{Csikor:2002ic}
\bibitem{Csikor:2002ic} 
  F.~Csikor, G.~I.~Egri, Z.~Fodor, S.~D.~Katz, K.~K.~Szabo and A.~I.~Toth,
  %``The QCD equation of state at finite T and mu,''
  Nucl.\ Phys.\ Proc.\ Suppl.\  {\bf 119}, 547 (2003)
  %doi:10.1016/S0920-5632(03)80453-X
  [hep-lat/0209114].
  %%CITATION = doi:10.1016/S0920-5632(03)80453-X;%%
  %12 citations counted in INSPIRE as of 07 Jul 2017


%\cite{Allton:2002zi}
\bibitem{Allton:2002zi} 
  C.~R.~Allton, S.~Ejiri, S.~J.~Hands, O.~Kaczmarek, F.~Karsch, E.~Laermann, C.~Schmidt and L.~Scorzato,
  %``The QCD thermal phase transition in the presence of a small chemical potential,''
  Phys.\ Rev.\ D {\bf 66}, 074507 (2002)
  %doi:10.1103/PhysRevD.66.074507
  [hep-lat/0204010].
  %%CITATION = doi:10.1103/PhysRevD.66.074507;%%
  %549 citations counted in INSPIRE as of 07 Jul 2017


%\cite{Allton:2005gk}
\bibitem{Allton:2005gk} 
  C.~R.~Allton, M.~Doring, S.~Ejiri, S.~J.~Hands, O.~Kaczmarek, F.~Karsch, E.~Laermann and K.~Redlich,
  %``Thermodynamics of two flavor QCD to sixth order in quark chemical potential,''
  Phys.\ Rev.\ D {\bf 71}, 054508 (2005)
  %doi:10.1103/PhysRevD.71.054508
  [hep-lat/0501030].
  %%CITATION = doi:10.1103/PhysRevD.71.054508;%%
  %422 citations counted in INSPIRE as of 07 Jul 2017


%\cite{Gavai:2008zr}
\bibitem{Gavai:2008zr} 
  R.~V.~Gavai and S.~Gupta,
  %``QCD at finite chemical potential with six time slices,''
  Phys.\ Rev.\ D {\bf 78}, 114503 (2008)
  %doi:10.1103/PhysRevD.78.114503
  [arXiv:0806.2233 [hep-lat]].
  %%CITATION = doi:10.1103/PhysRevD.78.114503;%%
  %194 citations counted in INSPIRE as of 07 Jul 2017


%\cite{Basak:2009uv}
\bibitem{Basak:2009uv} 
  S.~Basak {\it et al.} [MILC Collaboration],
  %``QCD equation of state at non-zero chemical potential,''
  PoS LATTICE {\bf 2008}, 171 (2008)
  [arXiv:0910.0276 [hep-lat]].
  %%CITATION = ARXIV:0910.0276;%%
  %12 citations counted in INSPIRE as of 07 Jul 2017


%\cite{deForcrand:2002hgr}
\bibitem{deForcrand:2002hgr} 
  P.~de Forcrand and O.~Philipsen,
  %``The QCD phase diagram for small densities from imaginary chemical potential,''
  Nucl.\ Phys.\ B {\bf 642}, 290 (2002)
  %doi:10.1016/S0550-3213(02)00626-0
  [hep-lat/0205016].
  %%CITATION = doi:10.1016/S0550-3213(02)00626-0;%%
  %608 citations counted in INSPIRE as of 07 Jul 2017


%\cite{DElia:2002tig}
\bibitem{DElia:2002tig} 
  M.~D'Elia and M.~P.~Lombardo,
  %``Finite density QCD via imaginary chemical potential,''
  Phys.\ Rev.\ D {\bf 67}, 014505 (2003)
  %doi:10.1103/PhysRevD.67.014505
  [hep-lat/0209146].
  %%CITATION = doi:10.1103/PhysRevD.67.014505;%%
  %473 citations counted in INSPIRE as of 07 Jul 2017


%\cite{Wu:2006su}
\bibitem{Wu:2006su} 
  L.~K.~Wu, X.~Q.~Luo and H.~S.~Chen,
  %``Phase structure of lattice QCD with two flavors of Wilson quarks at finite temperature and chemical potential,''
  Phys.\ Rev.\ D {\bf 76}, 034505 (2007)
  %doi:10.1103/PhysRevD.76.034505
  [hep-lat/0611035].
  %%CITATION = doi:10.1103/PhysRevD.76.034505;%%
  %89 citations counted in INSPIRE as of 07 Jul 2017


%\cite{deForcrand:2008vr}
\bibitem{deForcrand:2008vr} 
  P.~de Forcrand and O.~Philipsen,
  %``The Chiral critical point of N(f) = 3 QCD at finite density to the order (mu/T)**4,''
  JHEP {\bf 0811}, 012 (2008)
  %doi:10.1088/1126-6708/2008/11/012
  [arXiv:0808.1096 [hep-lat]].
  %%CITATION = doi:10.1088/1126-6708/2008/11/012;%%
  %139 citations counted in INSPIRE as of 07 Jul 2017


%\cite{DElia:2009pdy}
\bibitem{DElia:2009pdy} 
  M.~D'Elia and F.~Sanfilippo,
  %``Thermodynamics of two flavor QCD from imaginary chemical potentials,''
  Phys.\ Rev.\ D {\bf 80}, 014502 (2009)
  %doi:10.1103/PhysRevD.80.014502
  [arXiv:0904.1400 [hep-lat]].
  %%CITATION = doi:10.1103/PhysRevD.80.014502;%%
  %68 citations counted in INSPIRE as of 07 Jul 2017


%\cite{Philipsen:2014rpa}
\bibitem{Philipsen:2014rpa} 
  O.~Philipsen and C.~Pinke,
  %``Nature of the Roberge-Weiss transition in $N_f=2$ QCD with Wilson fermions,''
  Phys.\ Rev.\ D {\bf 89}, 094504 (2014)
  %doi:10.1103/PhysRevD.89.094504
  [arXiv:1402.0838 [hep-lat]].
  %%CITATION = doi:10.1103/PhysRevD.89.094504;%%
  %16 citations counted in INSPIRE as of 07 Jul 2017


%\cite{Cuteri:2015qkq}
\bibitem{Cuteri:2015qkq} 
  C.~Czaban, F.~Cuteri, O.~Philipsen, C.~Pinke and A.~Sciarra,
  %``Roberge-Weiss transition in $N_\text{f}=2$ QCD with Wilson fermions and $N_\tau=6$,''
  Phys.\ Rev.\ D {\bf 93}, 054507 (2016)
  %doi:10.1103/PhysRevD.93.054507
  [arXiv:1512.07180 [hep-lat]].
  %%CITATION = doi:10.1103/PhysRevD.93.054507;%%
  %5 citations counted in INSPIRE as of 07 Jul 2017

%\cite{DElia:2016jqh}
\bibitem{DElia:2016jqh} 
  M.~D'Elia, G.~Gagliardi and F.~Sanfilippo,
  %``Higher order quark number fluctuations via imaginary chemical potentials in $N_f=2+1$ QCD,''
  Phys.\ Rev.\ D {\bf 95}, 094503 (2017)
  %doi:10.1103/PhysRevD.95.094503
  [arXiv:1611.08285 [hep-lat]].
  %%CITATION = doi:10.1103/PhysRevD.95.094503;%%
  %13 citations counted in INSPIRE as of 26 Jul 2017

%\cite{Alba:2017mqu}
\bibitem{Alba:2017mqu} 
  P.~Alba {\it et al.},
  %``Constraining the hadronic spectrum through QCD thermodynamics on the lattice,''
  arXiv:1702.01113 [hep-lat].
  %%CITATION = ARXIV:1702.01113;%%
  %3 citations counted in INSPIRE as of 07 Jul 2017


%\cite{Bellwied:2015rza}
\bibitem{Bellwied:2015rza} 
  R.~Bellwied, S.~Borsanyi, Z.~Fodor, J.~Günther, S.~D.~Katz, C.~Ratti and K.~K.~Szabo,
  %``The QCD phase diagram from analytic continuation,''
  Phys.\ Lett.\ B {\bf 751}, 559 (2015)
  %doi:10.1016/j.physletb.2015.11.011
  [arXiv:1507.07510 [hep-lat]].
  %%CITATION = doi:10.1016/j.physletb.2015.11.011;%%
  %43 citations counted in INSPIRE as of 07 Jul 2017


%\cite{Gunther:2016vcp}
\bibitem{Gunther:2016vcp} 
  J.~Gunther, R.~Bellwied, S.~Borsanyi, Z.~Fodor, S.~D.~Katz, A.~Pasztor and C.~Ratti,
  %``The QCD equation of state at finite density from analytical continuation,''
  %EPJ Web Conf.\  {\bf 137}, 07008 (2017)
  %doi:10.1051/epjconf/201713707008
  [arXiv:1607.02493 [hep-lat]].
  %%CITATION = doi:10.1051/epjconf/201713707008;%%
  %23 citations counted in INSPIRE as of 07 Jul 2017

%\cite{Bazavov:2017dus}
\bibitem{Bazavov:2017dus} 
  A.~Bazavov {\it et al.},
  %``The QCD Equation of State to $\mathcal{O}(\mu_B^6)$ from Lattice QCD,''
  Phys.\ Rev.\ D {\bf 95},  054504 (2017)
  %doi:10.1103/PhysRevD.95.054504
  [arXiv:1701.04325 [hep-lat]].
  %%CITATION = doi:10.1103/PhysRevD.95.054504;%%
  %19 citations counted in INSPIRE as of 07 Jul 2017


%\cite{Borsanyi:QM2017}
\bibitem{Borsanyi:QM2017} 
  S.~Bors\'anyi {\it et al.} [Wuppertal-Budapest Collaboration],
  \href{https://indico.cern.ch/event/433345/contributions/2358551/attachments/1408969/2154260/qm17_borsanyi.pdf}{Talk} at Quark Matter 2017 conference (5-11 February 2017, Chicago, USA). 

%\cite{Bluhm:2007cp}
\bibitem{Bluhm:2007cp} 
  M.~Bluhm and B.~Kampfer,
  %``Quasiparticle model of quark-gluon plasma at imaginary chemical potential,''
  Phys.\ Rev.\ D {\bf 77}, 034004 (2008)
  %doi:10.1103/PhysRevD.77.034004
  [arXiv:0711.0590 [hep-ph]].
  %%CITATION = doi:10.1103/PhysRevD.77.034004;%%
  %29 citations counted in INSPIRE as of 12 Jul 2017

%\cite{Morita:2011jva}
\bibitem{Morita:2011jva} 
  K.~Morita, V.~Skokov, B.~Friman and K.~Redlich,
  %``Role of mesonic fluctuations in the Polyakov loop extended quark-meson model at imaginary chemical potential,''
  Phys.\ Rev.\ D {\bf 84}, 074020 (2011)
  %doi:10.1103/PhysRevD.84.074020
  [arXiv:1108.0735 [hep-ph]].
  %%CITATION = doi:10.1103/PhysRevD.84.074020;%%
  %17 citations counted in INSPIRE as of 12 Jul 2017

%\cite{Dashen:1969ep}
\bibitem{Dashen:1969ep} 
  R.~Dashen, S.~K.~Ma and H.~J.~Bernstein,
  %``S Matrix formulation of statistical mechanics,''
  Phys.\ Rev.\  {\bf 187}, 345 (1969).
  %doi:10.1103/PhysRev.187.345
  %%CITATION = doi:10.1103/PhysRev.187.345;%%
  %270 citations counted in INSPIRE as of 07 Jul 2017


%\cite{Cleymans:1992zc}
\bibitem{Cleymans:1992zc} 
  J.~Cleymans and H.~Satz,
  %``Thermal hadron production in high-energy heavy ion collisions,''
  Z.\ Phys.\ C {\bf 57}, 135 (1993)
  %doi:10.1007/BF01555746
  [hep-ph/9207204].
  %%CITATION = doi:10.1007/BF01555746;%%
  %252 citations counted in INSPIRE as of 07 Jul 2017


%\cite{Cleymans:1998fq}
\bibitem{Cleymans:1998fq} 
  J.~Cleymans and K.~Redlich,
  %``Unified description of freezeout parameters in relativistic heavy ion collisions,''
  Phys.\ Rev.\ Lett.\  {\bf 81}, 5284 (1998)
  %doi:10.1103/PhysRevLett.81.5284
  [nucl-th/9808030].
  %%CITATION = doi:10.1103/PhysRevLett.81.5284;%%
  %346 citations counted in INSPIRE as of 07 Jul 2017


%\cite{Becattini:2003wp}
\bibitem{Becattini:2003wp} 
  F.~Becattini, M.~Gazdzicki, A.~Keranen, J.~Manninen and R.~Stock,
  %``Chemical equilibrium in nucleus nucleus collisions at relativistic energies,''
  Phys.\ Rev.\ C {\bf 69}, 024905 (2004)
  %doi:10.1103/PhysRevC.69.024905
  [hep-ph/0310049].
  %%CITATION = doi:10.1103/PhysRevC.69.024905;%%
  %269 citations counted in INSPIRE as of 07 Jul 2017


%\cite{Andronic:2005yp}
\bibitem{Andronic:2005yp} 
  A.~Andronic, P.~Braun-Munzinger and J.~Stachel,
  %``Hadron production in central nucleus-nucleus collisions at chemical freeze-out,''
  Nucl.\ Phys.\ A {\bf 772}, 167 (2006)
  %doi:10.1016/j.nuclphysa.2006.03.012
  [nucl-th/0511071].
  %%CITATION = doi:10.1016/j.nuclphysa.2006.03.012;%%
  %557 citations counted in INSPIRE as of 07 Jul 2017


%\cite{Letessier:2005qe}
\bibitem{Letessier:2005qe} 
  J.~Letessier and J.~Rafelski,
  %``Hadron production and phase changes in relativistic heavy ion collisions,''
  Eur.\ Phys.\ J.\ A {\bf 35}, 221 (2008)
  %doi:10.1140/epja/i2007-10546-7
  [nucl-th/0504028].
  %%CITATION = doi:10.1140/epja/i2007-10546-7;%%
  %106 citations counted in INSPIRE as of 07 Jul 2017


%\cite{Olive:2016xmw}
\bibitem{Olive:2016xmw} 
  C.~Patrignani {\it et al.} [Particle Data Group],
  %``Review of Particle Physics,''
  Chin.\ Phys.\ C {\bf 40},  100001 (2016).
  %doi:10.1088/1674-1137/40/10/100001
  %%CITATION = doi:10.1088/1674-1137/40/10/100001;%%
  %1139 citations counted in INSPIRE as of 07 Jul 2017


%\cite{Becattini:1995if}
\bibitem{Becattini:1995if} 
  F.~Becattini,
  %``A Thermodynamical approach to hadron production in e+ e- collisions,''
  Z.\ Phys.\ C {\bf 69},  485 (1996).
  %doi:10.1007/BF02907431
  %%CITATION = doi:10.1007/BF02907431;%%
  %321 citations counted in INSPIRE as of 07 Jul 2017


%\cite{Wheaton:2004qb}
\bibitem{Wheaton:2004qb} 
  S.~Wheaton, J.~Cleymans and M.Hauer,
  %``THERMUS: A Thermal model package for ROOT,''
  Comput.\ Phys.\ Commun.\  {\bf 180}, 84 (2009)
  %doi:10.1016/j.cpc.2008.08.001
  [hep-ph/0407174].
  %%CITATION = doi:10.1016/j.cpc.2008.08.001;%%
  %230 citations counted in INSPIRE as of 07 Jul 2017

%\cite{Satarov:2016peb}
\bibitem{Satarov:2016peb} 
  L.~M.~Satarov, V.~Vovchenko, P.~Alba, M.~I.~Gorenstein and H.~Stoecker,
  %``New scenarios for hard-core interactions in a hadron resonance gas,''
  Phys.\ Rev.\ C {\bf 95}, 024902 (2017)
  %doi:10.1103/PhysRevC.95.024902
  [arXiv:1610.08753 [nucl-th]].
  %%CITATION = doi:10.1103/PhysRevC.95.024902;%%
  %7 citations counted in INSPIRE as of 08 Aug 2017

%\cite{Rischke:1991ke}
\bibitem{Rischke:1991ke} 
  D.~H.~Rischke, M.~I.~Gorenstein, H.~Stoecker and W.~Greiner,
  %``Excluded volume effect for the nuclear matter equation of state,''
  Z.\ Phys.\ C {\bf 51}, 485 (1991).
  %doi:10.1007/BF01548574
  %%CITATION = doi:10.1007/BF01548574;%%
  %190 citations counted in INSPIRE as of 07 Jul 2017


%\cite{BraunMunzinger:1999qy}
\bibitem{BraunMunzinger:1999qy} 
  P.~Braun-Munzinger, I.~Heppe and J.~Stachel,
  %``Chemical equilibration in Pb + Pb collisions at the SPS,''
  Phys.\ Lett.\ B {\bf 465}, 15 (1999)
  %doi:10.1016/S0370-2693(99)01076-X
  [nucl-th/9903010].
  %%CITATION = doi:10.1016/S0370-2693(99)01076-X;%%
  %566 citations counted in INSPIRE as of 07 Jul 2017


%\cite{Cleymans:2005xv}
\bibitem{Cleymans:2005xv} 
  J.~Cleymans, H.~Oeschler, K.~Redlich and S.~Wheaton,
  %``Comparison of chemical freeze-out criteria in heavy-ion collisions,''
  Phys.\ Rev.\ C {\bf 73}, 034905 (2006)
  %doi:10.1103/PhysRevC.73.034905
  [hep-ph/0511094].
  %%CITATION = doi:10.1103/PhysRevC.73.034905;%%
  %402 citations counted in INSPIRE as of 07 Jul 2017


%\cite{Randrup:2009ch}
\bibitem{Randrup:2009ch} 
  J.~Randrup and J.~Cleymans,
  %``Exploring high-density baryonic matter: Maximum freeze-out density,''
  Eur.\ Phys.\ J.\  {\bf 52}, 218 (2016)
  %doi:10.1140/epja/i2016-16218-7
  [arXiv:0905.2824 [nucl-th]].
  %%CITATION = doi:10.1140/epja/i2016-16218-7;%%
  %11 citations counted in INSPIRE as of 07 Jul 2017


%\cite{Andronic:2012ut}
\bibitem{Andronic:2012ut} 
  A.~Andronic, P.~Braun-Munzinger, J.~Stachel and M.~Winn,
  %``Interacting hadron resonance gas meets lattice QCD,''
  Phys.\ Lett.\ B {\bf 718}, 80 (2012)
  %doi:10.1016/j.physletb.2012.10.001
  [arXiv:1201.0693 [nucl-th]].
  %%CITATION = doi:10.1016/j.physletb.2012.10.001;%%
  %59 citations counted in INSPIRE as of 07 Jul 2017


%\cite{Vovchenko:2014pka}
\bibitem{Vovchenko:2014pka} 
  V.~Vovchenko, D.~V.~Anchishkin and M.~I.~Gorenstein,
  %``Hadron Resonance Gas Equation of State from Lattice QCD,''
  Phys.\ Rev.\ C {\bf 91}, 024905 (2015)
  %doi:10.1103/PhysRevC.91.024905
  [arXiv:1412.5478 [nucl-th]].
  %%CITATION = doi:10.1103/PhysRevC.91.024905;%%
  %24 citations counted in INSPIRE as of 07 Jul 2017

%\cite{Alba:2016fku}
\bibitem{Alba:2016fku} 
  P.~Alba, W.~M.~Alberico, A.~Nada, M.~Panero and H.~St\"ocker,
  %``Excluded-volume effects for a hadron gas in Yang-Mills theory,''
  Phys.\ Rev.\ D {\bf 95}, 094511 (2017)
  %doi:10.1103/PhysRevD.95.094511
  [arXiv:1611.05872 [hep-lat]].
  %%CITATION = doi:10.1103/PhysRevD.95.094511;%%
  %2 citations counted in INSPIRE as of 18 Aug 2017

%\cite{Roberge:1986mm}
\bibitem{Roberge:1986mm} 
  A.~Roberge and N.~Weiss,
  %``Gauge Theories With Imaginary Chemical Potential and the Phases of {QCD},''
  Nucl.\ Phys.\ B {\bf 275}, 734 (1986).
  %doi:10.1016/0550-3213(86)90582-1
  %%CITATION = doi:10.1016/0550-3213(86)90582-1;%%
  %297 citations counted in INSPIRE as of 26 Jul 2017

%\cite{Borsanyi:2016ksw}
\bibitem{Borsanyi:2016ksw} 
  S.~Borsanyi {\it et al.},
  %``Calculation of the axion mass based on high-temperature lattice quantum chromodynamics,''
  Nature {\bf 539}, 69 (2016)
  %doi:10.1038/nature20115
  [arXiv:1606.07494 [hep-lat]].
  %%CITATION = doi:10.1038/nature20115;%%
  %61 citations counted in INSPIRE as of 27 Jul 2017

%\cite{Danzer:2012vw}
\bibitem{Danzer:2012vw} 
  J.~Danzer and C.~Gattringer,
  %``Properties of canonical determinants and a test of fugacity expansion for finite density lattice QCD with Wilson fermions,''
  Phys.\ Rev.\ D {\bf 86}, 014502 (2012)
  %doi:10.1103/PhysRevD.86.014502
  [arXiv:1204.1020 [hep-lat]].
  %%CITATION = doi:10.1103/PhysRevD.86.014502;%%
  %25 citations counted in INSPIRE as of 25 Sep 2017

%\cite{Gattringer:2014hra}
\bibitem{Gattringer:2014hra} 
  C.~Gattringer and H.~P.~Schadler,
  %``Generalized quark number susceptibilities from fugacity expansion at finite chemical potential for $N_f$ = 2 Wilson fermions,''
  Phys.\ Rev.\ D {\bf 91}, no. 7, 074511 (2015)
  %doi:10.1103/PhysRevD.91.074511
  [arXiv:1411.5133 [hep-lat]].
  %%CITATION = doi:10.1103/PhysRevD.91.074511;%%
  %16 citations counted in INSPIRE as of 25 Sep 2017

%\cite{Huovinen:2017ogf}
\bibitem{Huovinen:2017ogf}
    P.~Huovinen and P.~Petreczky,
    %``Hadron Resonance Gas with Repulsive Interactions and Fluctuations of Conserved Charges,''
    arXiv:1708.00879 [hep-ph].
    %%CITATION = ARXIV:1708.00879;%%

%\cite{DElia:2007bkz}
\bibitem{DElia:2007bkz} 
  M.~D'Elia, F.~Di Renzo and M.~P.~Lombardo,
  %``The Strongly interacting quark gluon plasma, and the critical behaviour of QCD at imaginary mu,''
  Phys.\ Rev.\ D {\bf 76}, 114509 (2007)
  %doi:10.1103/PhysRevD.76.114509
  [arXiv:0705.3814 [hep-lat]].
  %%CITATION = doi:10.1103/PhysRevD.76.114509;%%
  %89 citations counted in INSPIRE as of 16 Sep 2017

%\cite{Takahashi:2014rta}
\bibitem{Takahashi:2014rta} 
  J.~Takahashi, H.~Kouno and M.~Yahiro,
  %``Quark number densities at imaginary chemical potential in $N_f=2$ lattice QCD with Wilson fermions and its model analyses,''
  Phys.\ Rev.\ D {\bf 91}, no. 1, 014501 (2015)
  %doi:10.1103/PhysRevD.91.014501
  [arXiv:1410.7518 [hep-lat]].
  %%CITATION = doi:10.1103/PhysRevD.91.014501;%%
  %17 citations counted in INSPIRE as of 16 Sep 2017

%\cite{Bornyakov:2016wld}
\bibitem{Bornyakov:2016wld} 
  V.~G.~Bornyakov, D.~L.~Boyda, V.~A.~Goy, A.~V.~Molochkov, A.~Nakamura, A.~A.~Nikolaev and V.~I.~Zakharov,
  %``New approach to canonical partition functions computation in $N_f=2$ lattice QCD at finite baryon density,''
  Phys.\ Rev.\ D {\bf 95}, no. 9, 094506 (2017)
  %doi:10.1103/PhysRevD.95.094506
  [arXiv:1611.04229 [hep-lat]].
  %%CITATION = doi:10.1103/PhysRevD.95.094506;%%
  %4 citations counted in INSPIRE as of 16 Sep 2017

%\cite{Boyda:2017dyo}
\bibitem{Boyda:2017dyo} 
  D.~Boyda, V.~G.~Bornyakov, V.~Goy, A.~Molochkov, A.~Nakamura, A.~Nikolaev and V.~I.~Zakharov,
  %``Lattice Study of QCD Phase Structure by Canonical Approach,''
  arXiv:1704.03980 [hep-lat].
  %%CITATION = ARXIV:1704.03980;%%

%\cite{Beth:1937zz}
\bibitem{Beth:1937zz} 
  E.~Beth and G.~Uhlenbeck,
  %``The quantum theory of the non-ideal gas. II. Behaviour at low temperatures,''
  Physica {\bf 4}, 915 (1937).
  %doi:10.1016/S0031-8914(37)80189-5
  %%CITATION = doi:10.1016/S0031-8914(37)80189-5;%%
  %178 citations counted in INSPIRE as of 27 Jul 2017

%\cite{Bazavov:2017tot}
\bibitem{Bazavov:2017tot} 
  A.~Bazavov {\it et al.} [HotQCD Collaboration],
  %``Skewness and kurtosis of net baryon-number distributions at small values of the baryon chemical potential,''
  arXiv:1708.04897 [hep-lat].
  %%CITATION = ARXIV:1708.04897;%%
  %1 citations counted in INSPIRE as of 29 Sep 2017

%\cite{Bazavov:2013dta}
\bibitem{Bazavov:2013dta} 
  A.~Bazavov {\it et al.},
  %``Strangeness at high temperatures: from hadrons to quarks,''
  Phys.\ Rev.\ Lett.\  {\bf 111}, 082301 (2013)
  %doi:10.1103/PhysRevLett.111.082301
  [arXiv:1304.7220 [hep-lat]].
  %%CITATION = doi:10.1103/PhysRevLett.111.082301;%%
  %92 citations counted in INSPIRE as of 13 Sep 2017

%\cite{Vovchenko:2017zpj}
\bibitem{Vovchenko:2017zpj} 
  V.~Vovchenko, A.~Motornenko, P.~Alba, M.~I.~Gorenstein, L.~M.~Satarov and H.~Stoecker,
  %``Multi-component van der Waals equation of state: applications in nuclear and hadronic physics,''
  arXiv:1707.09215 [nucl-th].
  %%CITATION = ARXIV:1707.09215;%%
  %1 citations counted in INSPIRE as of 05 Aug 2017

\bibitem{juqueen}
  JUQUEEN: IBM Blue Gene/Q Supercomputer System at the J\"ulich Supercomputing Centre,
  Tech. Rep. 1 A1 (J\"ulich Supercomputing Centre, http://dx.doi.org/10.17815/jlsrf-1-18, 2015).


\end{thebibliography}
\end{document}